\documentclass{webofc}
\usepackage[varg]{txfonts}   % Web of Conferences font
\usepackage{float}
\usepackage{placeins}

\usepackage{hyperref}
\usepackage{etoolbox}
\AtEndDocument{{\linespread{0.9}\bibliography{biblio,physics}}}

\usepackage[breakable]{tcolorbox}
\newtcolorbox{templatebox}[1][]{breakable,#1}
\usepackage{subfiles}

\usepackage[textsize=tiny]{todonotes}
\newcounter{todocounter}

\usepackage{lineno}

\usepackage{relsize}
\usepackage{xspace}

\usepackage{graphicx}
\usepackage[labelfont=bf,labelsep=period]{caption}
\usepackage{subcaption}

\newcommand\cpp{C\nolinebreak[4]\hspace{-.05em}\raisebox{.2ex}{\relscale{1}{++}}\xspace}
\newcommand\typename[1]{\texttt{#1}}
\newcommand\Bundle{\typename{Bundle}\xspace}

\newcommand\DataType{\typename{Descriptor}\xspace}

\newcommand\Expression{\typename{Expression}\xspace}

\newlength{\figureA}
\newlength{\figureB}
\newlength{\figureC}
\begin{document}

\title{GNA: new framework for statistical data analysis}


\author{\firstname{Anna}~\lastname{Fatkina}\inst{1} \and
        \firstname{Maxim}~\lastname{Gonchar}\inst{1}\fnsep\thanks{\email{gonchar@jinr.ru}} \and
        \firstname{Anastasia}~\lastname{Kalitkina}\inst{1} \and
        \firstname{Liudmila}~\lastname{Kolupaeva}\inst{1} \and
        \firstname{Dmitry}~\lastname{Naumov}\inst{1} \and
        \firstname{Dmitry}~\lastname{Selivanov}\inst{1} \and
        \firstname{Konstantin}~\lastname{Treskov}\inst{1}
}

\institute{Joint Institute for Nuclear Research, Dubna, Moscow, Russia}

\abstract{%
  We report on the status of GNA --- a new framework for fitting large-scale physical models. GNA utilizes the data flow
  concept within which a model is represented by a directed acyclic graph. Each node is an operation on an array (matrix
  multiplication, derivative or cross section calculation, etc). The framework enables the user to create flexible and
  efficient large-scale lazily evaluated models, handle large numbers of parameters, propagate parameters' uncertainties while
  taking into account possible correlations between them, fit models, and perform statistical analysis.

  The main goal of the paper is to give an overview of the main concepts and methods as well as reasons behind their design. 
  Detailed technical information is to be published in further works.
}

\maketitle

%\listoftodos
%\tableofcontents

\section{Introduction}
\subsection{Interpretation of observables}

The measurement of physical parameters from experimental data
significantly relies on methods of statistical inference. The main
methods of performing parameter inference are either maximum-likelihood-based (ML) or Bayesian.
These approaches involve massive numerical computations when working with
models with a large number of parameters, either free or constrained.

Searching for evidence of weak signals or having a limited sample of
data often requires special numerical procedures to derive confidence
intervals for parameters of interest. Examples of
such approaches are Feldman-Cousins (FC)~\cite{1998PhRvD..57.3873F} (pure frequentist) and
Cousins-Highland \cite{Cousins:1991qz} (hybrid Bayesian-frequentist) methods.
The guarantee of correct coverage provided by these techniques comes at the
cost of extensive numerical computations, e.g. fitting a model to a large number
of pseudo-experiments to derive the test statistic distribution.

Both signal weakness and finiteness of the data sample are common situations in modern
neutrino physics.
Searches for a light sterile neutrino in Daya Bay~\cite{An:2016luf} and oscillation analysis of
NOvA~\cite{Adamson:2016tbq} are two examples among many of this kind, that use methods requiring high performance
computing.

Performing high load statistical analysis with large-scale models requires dedicated software.
Extra attention should be paid to long-term analysis consistency since the typical lifetime
of a modern neutrino physics experiment may achieve dozens of years.

\subsection{Software}

The most common tool to perform statistical data analysis used in high
energy physics is currently ROOT~\cite{Brun:1997pa} with the RooFit package~\cite{Verkerke:2003ir}
built on top of it.
These tools are well-suited for inference and parameter estimation for
relatively low-dimension problems with small numbers of model
parameters. When the number of parameters in the analysis grows these tools
become less flexible and efficient.
%Another
%drawback is that they don't benefit and scale with modern
%computational architectures such as GPGPUs and multithreading capabilities
%of currents CPUs.

A new generation of software frameworks taking advantage of modern
hardware and suitable for contemporary analysis needs is required.
A few solutions are already available.

GooFit \cite{Andreassen:2014xfa} is a library designed to perform
maximum-likelihood fits in a fashion similar to RooFit but taking
advantage of massive parallel computations with GPUs and modern CPUs
through NVIDIA Thrust and OpenMP. It is used extensively in LHCb
analyses~\cite{Aaij:2015lsa}.

Hydra \cite{hydra} is a header-only library for statistical data analysis
utilizing CUDA, OpenMP and TBB as backends. It offers
maximum-likelihood fitting, tools for working with multidimensional PDFs,
multidimensional numerical integration with MCMC, and other tools.

Both of these libraries require model specification at compile time, which
limits flexibility when working with varying models.

\section{GNA}
\subsection{Design principles}

Below we will list the typical analysis context and the features the user requires from the framework.

First of all we are aiming at large-scale models, such as Daya Bay model (variant D)~\cite{An:2016ses}, that may
yield several histograms with hundreds of bins, contain thousands of elements and depend on hundreds of model
parameters, some of them may be partially correlated. The parameters do
affect various parts of the model and may be varied one-by-one, group-by-group, or all together. The common
tasks done with a model include:
\begin{itemize}\itemsep=-4pt
  \item computation of the derivative over a parameter,
  \item multidimensional ML fit with few parameters or with with hundreds of parameters,
  \item 2d scan over two parameters with fit of all the remaining parameters in each point,
  \item numerous fits to MC data, e.g. hundreds of thousands of fits within FC method~\cite{1998PhRvD..57.3873F},
  \item combined study of a few models taking into account correlations between their parameters.
\end{itemize}

During the analysis preparation various parts of the model may be replaced by alternative parametrizations or approximations.
Even the computation scheme itself may be significantly refactored. After the analysis results are published the model
is maintained for a period of 5 to 10 years by different people with different experience in software development.
Each analysis update may again require a significant model update while the subsequent analyses should be consistent with
each other and the old analyses should be reproducible.

Based on this, the requirements to the framework include high flexibility, computational efficiency, expressiveness,
easy maintainability and long-term backwards compatibility.

We are solving these issues by using the data flow paradigm~\cite{DenniDataflow}. The model is build at runtime as a directed acyclic
graph. Each node represent an operation on an input data array(s) producing several output data arrays.
The framework provides a library of precompiled operations, also called transformations,
that are required for the statistical analysis of physical data.
The model is then assembled by binding instances of transformations together. This enables the
user to concentrate on the
logic and physics of the analysis while ignoring the low-level issues, such as memory allocation.
All of the internals are organized, accessible and reviewable. Computational graphs (chains) of different models may be
combined together.

The transformations and transformation bundles are designed to be immutable. Instead of modifying the transformation
it is suggested to extend the library by writing another one, which facilitates backwards compatibility.

%We implement an expression parser that enables user to define the computational graph structure by providing a
%mathematical expression, which makes GNA very flexible.

The computational chain, maintained by GNA, is in fact a runtime call graph.
This facilitates the implementation, implicit from the user's point of view, of
powerful features, such as CPU multithreading, GPU support, etc.

There are a lot of parallels and similarities with the TensorFlow package~\cite{tensorflow2015-whitepaper} as well as
some other data flow tools for neural networks. In fact, we are trying to simplify similar software tasks within a scope of
statistical analysis of large scale physical models.

\subsection{GNA core}

GNA is implemented in \cpp and Python. The core is implemented in \cpp to provide efficiency while the
management code is written in Python to provide flexibility. GNA depends on the Eigen~\cite{eigenweb} library for array
operations and ROOT package~\cite{Brun:1997pa} responsible for minimization and \cpp/Python binding. Other dependencies
include boost~\cite{boost}, GSL~\cite{DBLP:books/daglib/0026677}, numpy/scipy~\cite{scipy} and
matplotlib~\cite{Hunter:matplotlib}.

The model in GNA is represented by a computational graph, that consists of nodes (\typename{Transformation}) and edges
connecting transformations' outputs (\typename{Sink}) to another transformations' inputs (\typename{Source}). The data
arrays are allocated within transformation sinks. The sources, in contrast, hold no data: the
source is a view to another transformation sink.

There are two templates for data types in GNA, \typename{variable} and \typename{Data}. \typename{variable} is a wrapper
for an elementary data type, usually double. Variables represent model parameters and may be used in a process of
minimization, or derivative calculation, etc.
%The variables are instantiated and managed on a Python side. Transformations may
%depend on variables: in this case the variables are bound to transformations on their initialization.

The \typename{Data} is an array data type that `flows' within a graph. It is allocated on transformations' sinks and
holds a buffer. The \typename{Data} is
characterized by a \DataType that holds the information on size, dimension and type of the data.
\typename{Data} may be a regular array or a histogram. In the latter case the \DataType also carries an array of
bin edges.

Variables and transformations carry \typename{taintflag} that indicates the validity of the stored data. The flags are
connected in a graph. Whenever a variable is modified it becomes tainted. The taint flag is then propagated
to all the dependent objects, variables and transformations, invalidating a part of the computational graph. When the
output of the untainted transformation is accessed the cached result is returned. If the transformation is tainted
the transformation function is called in order to update the data. The function itself may read the other
transformations' data triggering a chain of updates. Finally, only the part of the graph that is tainted and that is
required to build the requested data is executed, other parts either return cached data or are not touched at all. This
effectively implements lazy evaluation of a graph.

%\typename

\subsection{GNA usage}

The transformation is defined by several \cpp functions: one for calculation and a few for type checking. The calculation
function may access transformation sources and sinks and is used to compute the result. It is executed if
the output of tainted transformation is accessed.

The type functions serve two purposes. First, they check that input descriptors are consistent between each
other and the transformation requirements. Second, they derive the output descriptors which are used by
the framework to allocate memory. Type functions are evaluated whenever two transformations are bound together, source-to-sink.
For example, in the case of a transformation implementing matrix multiplication the user binds two matrices to the
transformation sources: $N\times M$ and $M\times K$. When inputs are bound, the transformation's first type function
will check that the matrices are indeed multipliable or throw an exception otherwise. The second type function will tell
that the output should have dimensions $N\times K$ and the framework will allocate the relevant array at
the only transformation sink.

Thus, the two stages may be outlined in GNA model usage: the binding stage and calculation stage. At the first stage,
when the transformations are being instantiated and bound, the framework derives data types and allocates memory. The
variables are bound to the transformations on the first stage as well. While the first stage may not be that efficient it
is typically executed only once. At the second stage the model is repeatedly used for computation. Since the framework
guarantees that the inputs and outputs passed to the computation functions are of proper shape the corresponding checks
are omitted in the computation functions, which makes them more readable and efficient.

A special remark should be made on the integration process. The GNA work flow implies that the transformations operate
on arrays and that at the calculation stage the computational graph has all the required memory allocated. In such a
deterministic scheme the integration process should also be explicitly vectorized, which also means that the integration
precision is chosen beforehand. Since GNA is aimed at fitting models to data we assume that in most cases the
integration precision may indeed be defined in prior.

The integration process is then split into two transformations. For a given set of bin
edges $x_i$ and integration orders $M_i$ the first transformation samples all the points $p_{ij}$ and weights
$\omega_{ij}$ needed to compute the integral for each bin. The set of points $p_{ij}$ is then passed to a function of
interest\footnotemark{} as an array. The second transformation then convolves the output of the function $f(p_{ij})$ with provided
weights, producing a histogram. The 1D case reads as follows:
\footnotetext{the function may be represented by a single transformation or a small computational chain.}
\begin{linenomath}
  \addtolength{\abovedisplayskip}{-6pt}
  \addtolength{\belowdisplayskip}{-6pt}
  \begin{align}
    H_i = \int\limits_{x_i}^{x_{i+1}} f(x) dx \approx \sum_{j_i=1}^{M_i} \omega_{ij} f(p_{ij}).
  \end{align}
\end{linenomath}
The weights and points are computed once. The method of Gauss-Legendre (GL) quadratures is implemented in GNA for 1 and
2 dimensions with an easy extension to other methods.

A similar deterministic approach is applied to spline interpolation. Since the target points are known beforehand the
binary search is extracted to a separate transformation which is executed only once. One of the benefits of the approach
is that it enables effective usage of the framework's caching nature.
For example, in the following integral
$\int\sigma(E)P(E)S(E)dE$ the computationally expensive cross section $\sigma(E)$ might not depend on varying model
parameters and
will be evaluated only once at runtime, while the computationally cheaper probability $P(E)$ and spectrum shape $S(E)$
depend on the minimization parameters and will be evaluated on each execution, following the variation of relevant
parameters.

\subsection{Transformations, bundles and expressions}

The GNA framework provides a library of transformations required for building models and performing statistical analysis of
data. Some of the included transformations are basic operations (sum, product), linear algebra (Cholesky decomposition),
statistics ($\chi^2$,
covariance matrix and Poisson log-likelihood calculation), calculus (differentiation and integration), physics (neutrino
oscillations probability, inverse beta decay cross section), and detector effects (energy smearing, energy response
distortion).
There is a focus on reactor neutrino physics due to the fact that the project has grown from the analysis of
Daya Bay experimental data.

Even after the transformations are introduced, combining them may be quite a tedious job. To make it easier we
introduce bundles and expressions. A \Bundle is a Python class that reads a dictionary containing the configuration and initializes a set
of variables and instantiates and binds a set of transformations. The result of \Bundle execution is a small
computational chain.

A special tool is provided for the purpose of binding small computational graphs together to form a large chain. The
concept called \Expression is using mathematical expressions to describe connections between transformations and
bundles.
\Expression is initialized with the following information: a) a mathematical expression,
b) the definition of indices used in the expression, c) configurations for the bundles to be executed to provide expression
elements.

The expressions are parsed within a predefined python environment and are valid python expressions. The only distinction
is that we are using the \verb+'func| arg1, arg2, ...'+ notation for \verb+'func(arg1, arg2, ...)'+ in order to
improve the readability of the chain calls. Consider the following example:
\begin{linenomath}
\addtolength{\abovedisplayskip}{-6pt}
\addtolength{\belowdisplayskip}{-6pt}
\begin{equation}
\verb;sum[k]| func[k]| scale[i] * vec[j]() + offset();
\label{lst:expression}
\end{equation}
\end{linenomath}
The expressions are parsed without actual knowledge of what data and functions are: each name is considered to be an
index, variable or transformation output. Here \verb+vec[j]()+ is a transformation output providing an array. The bundle
that will provide \verb+vec+ will provide an output for each variant of index \verb+j+. \verb+scale[k]+ is a variable with
variants for \verb+k1+, \verb+k2+, etc. The multiplication \verb+scale[i] * vec[j]()+ will provide a transformation that
scales an array for each combination of indices of \verb+i+ and \verb+j+. Then the \verb+offset+ array will be added to
each result.

\verb+func+ is a function with one argument and one return value. The call operation means that the
output, associated with the argument, is bound to the input, associated with a function. As in the case of multiplication each
permutation of indices is taken into account. Since there is an output for each combination of \verb+i+ and \verb+j+
values and functions have several variants, represented by index \verb+k+, a call will produce an output for
each permutation of \verb+i+, \verb+j+ and \verb+k+ variants. Lastly the \verb+sum[k]+ will make a sum over index
\verb+k+ and provide the output for each of \verb+i+ and \verb+j+ variants.
Then, for each unknown name the \Expression finds a bundle
configuration and executes the corresponding bundle, passing the indices assigned to the name. The bundle should then
provide all the necessary inputs, outputs and variables. Once bundles are executed \Expression will bind the
transformations together.

\begin{figure}[htpb]
  \vspace{-\figureA}
  \centering
  \includegraphics[width=1\linewidth,trim=38 40 40 40,clip]{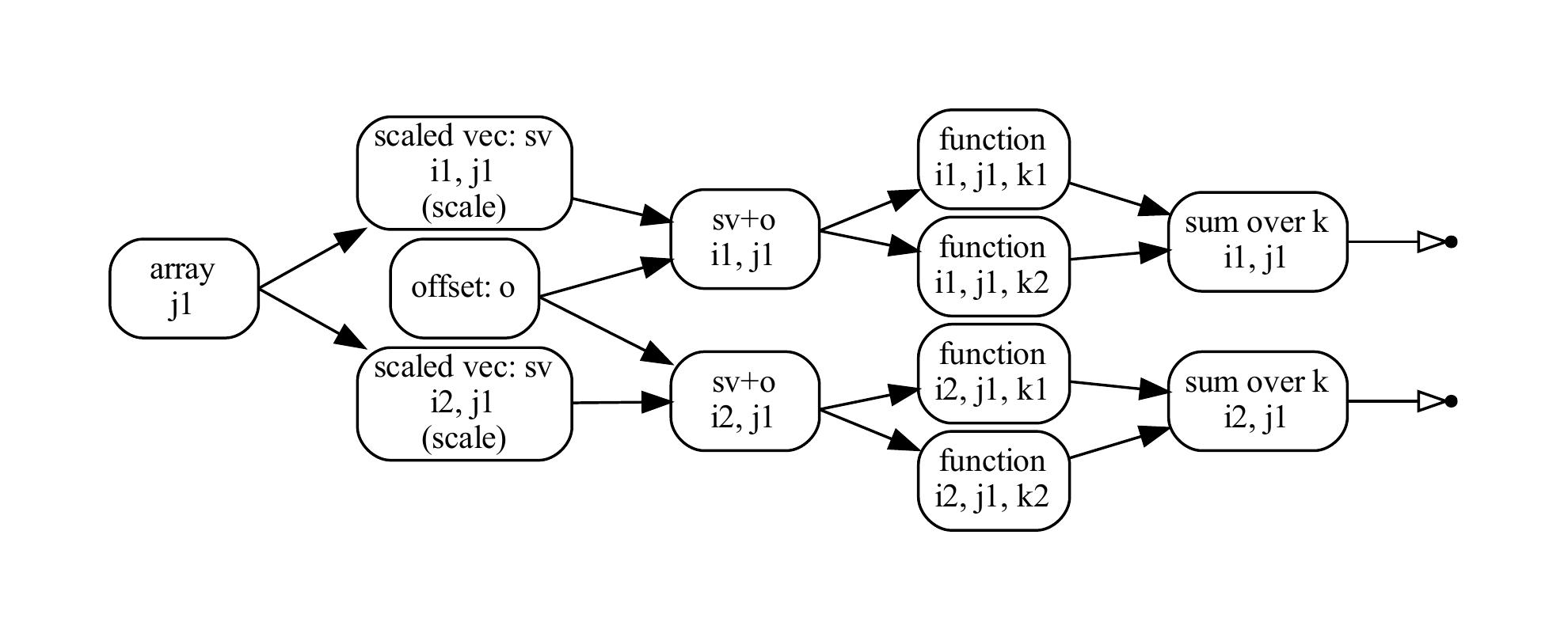}
  \vspace{-\figureC}
  \caption{An example of the computational chain, produced by expression~\eqref{lst:expression}.}
  \label{fig:sample_graph}
  \vspace{-\figureB}
\end{figure}

The result of the expression~\eqref{lst:expression} for the case when index \verb+i+ has variants
$\{\verb+i1+,\verb+i2+\}$, $\verb+j+\in\{\verb+j1+\}$ and $\verb+k+\in\{\verb+k1+,\verb+k2+\}$ is shown in
figure~\ref{fig:sample_graph}. The graph represents only transformations and their connections while variables are not
plotted.

The \Expression module provides operators for addition, multiplication and division; summation, multiplication and
concatenation over indices. These few elements, together with a set of bundles, is enough to build large-scale models. For
example, the Daya Bay model is described by 6 indices, roughly 50 items in the expression, 25 configuration items that are
using 16 bundles. The \Expression produces a computational chain of 732 nodes and 1624 edges (736 outputs with
13.6~Mb of data); 498 parameters (114 fixed, 97 variable and 287 evaluated). The full computational time is 10~ms.
Only a slight modification of the order of instructions in the expression yield a very different computational graph with
2412 nodes and 5968 edges (4416 outputs with 20.3 Mb of data). The latter chain produces identical result with a full
computational time of 20~ms. A more detailed description of the example may be found in~\cite{Gonchar:1810db}.

%\subsection{Other tools}
%\subfile{02e_tools.tex}
%\subsection{Current status}
%\subfile{02f_status.tex}

\section{Prospects}

Since GNA has full control over how the computational chain is executed it is possible to introduce
alternative mechanisms utilizing modern computation techniques on a level
of the framework, i.e. implicitly for the end user.
Our development team is working on GPU and multithreading support in order to make GNA
suitable for heterogeneous systems. In the following chapter an overview of the task manager for multicore CPU and GPU
support is given.

\subsection{Multithreading}

Different parts of the GNA computational graph may be executed in parallel threads with each transformation representing
a task. Due to the lazy and caching nature of GNA the evaluation order is not fixed and depends on the
computational chain state and the transformation which the output is called. The multithreading support was introduced in
order to manage execution and assign tasks to workers at runtime.

A thread pool is implemented with workers assigned to parallel threads.
When the transformation output is called the support system traverses the computational graph and pushes the tainted
transformations into workers' stacks as distinct tasks. The first subsequent transformation is pushed into the current thread worker's
stack. The other transformations are pushed into either new workers' stacks or free workers' stacks. In
case the task is accessing the transformation which is being computed in another thread the reading is blocked until the
computation is finished.

\begin{figure}[h]
  \vspace{-\figureA}
  \centering
  \begin{subfigure}[b]{0.55\textwidth}
    \includegraphics[width=0.95\textwidth]{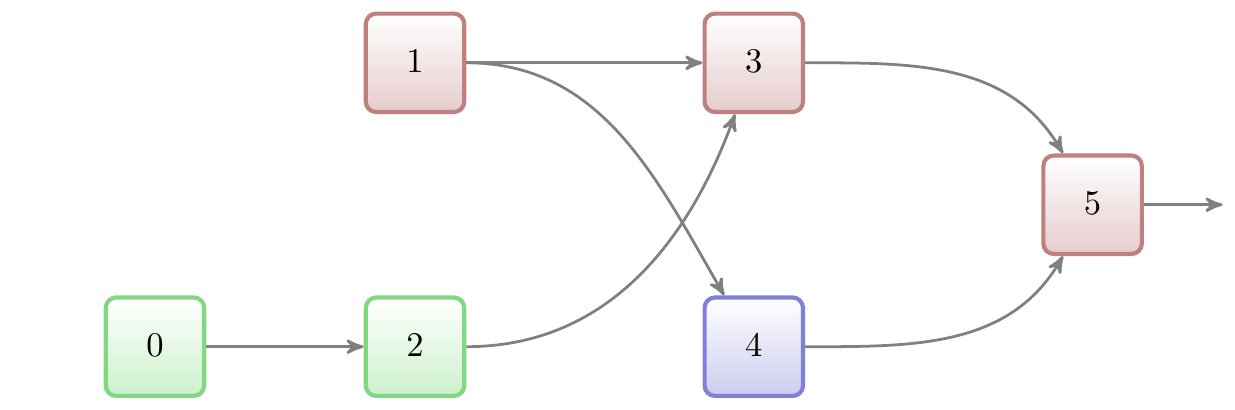}
    \caption{}
    \label{fig:workers_scheme}
    \vspace{-\figureC}
  \end{subfigure}
  \hfill
  \begin{subfigure}[b]{0.4\textwidth}
    \includegraphics[width=0.9\textwidth]{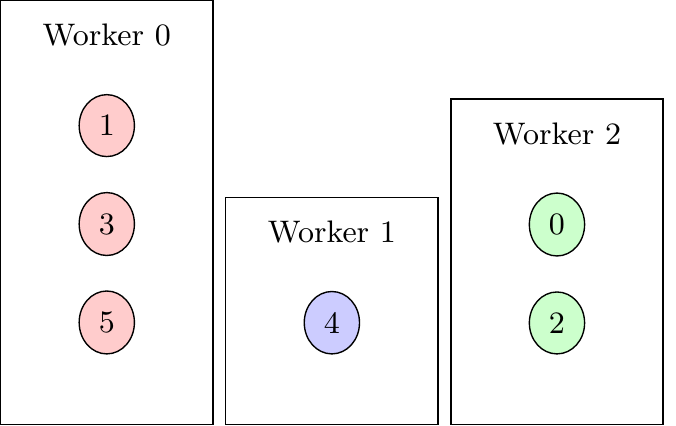}
    \caption{}
    \label{fig:workers_stack}
    \vspace{-\figureC}
  \end{subfigure}
  \vspace{-\figureC}
  \caption{Task management scheme for the case of three threads. Example graph is shown in~(\subref{fig:workers_scheme})
    with colors representing the worker. Workers' stacks are shown in~(\subref{fig:workers_stack}) for the case when the
  output of 5th transformation is called.}
  \label{fig:workers}
  \vspace{-\figureB}
\end{figure}
%\FloatBarrier

There exist various strategies on how to assign tasks to workers. We have chosen the `wide' strategy that tries to
execute maximally independent branches in parallel threads. An example scheme of the runtime task management for 3 threads is shown in
figure~\ref{fig:workers}. When the output of transformation 5  (${T_5}$) is read it is pushed to the worker 0 (${W_0}$),
${T_3}$ is then pushed to the same worker and ${T_4}$ to the new worker ${W_1}$. On the next step the inputs of ${T_3}$
and ${T_4}$ are considered: ${T_1}$ is pushed to ${W_0}$ (same worker with ${T_3}$), ${T_2}$ to the new worker ${W_2}$.
${T_4}$ then will wait for the ${T_1}$ to be finished by ${W_0}$. ${T_0}$ is added to ${W_2}$.

Due to the fact that transformations, which access the same outputs from parallel threads, are temporarily blocked the efficiency of such an
approach to multithreading will be below 100\% and will depend on the computational chain structure. The actual
efficiency will be studied in the future.

\subsection{GPU}

GPU support through CUDA~\cite{nvidia2007compute,nickolls2008scalable} for GNA was recently announced in
\cite{DBLP:conf/iccsa/FatkinaGKNT18}. Preliminary test results were presented showing x20 speed up of neutrino
oscillation probability calculations when compared to the single-core CPU
version. We used a laptop powered by NVIDIA GTX 970M and Intel Core i7-6700HQ.
These tests were done with double precision floating-point operations since, at the moment, this is the only
precision supported by the framework.

GNA provides great flexibility due to the fact that individual transformations, while being compiled independently from each
other, may be combined at runtime into any computational scheme. The GPU support follows a similar paradigm.
Unified interfaces are introduced to provide GPU-based implementations of the transformations in a similar way as
CPU-versions are done. The transformation may be switched between CPU and GPU implementations at runtime.
GPU memory is allocated when required and implicit data transfers between CPU and GPU occur lazily at runtime.

\begin{figure}[h]
  \vspace{-\figureA}
  \centering
  \includegraphics[width=0.7\textwidth]{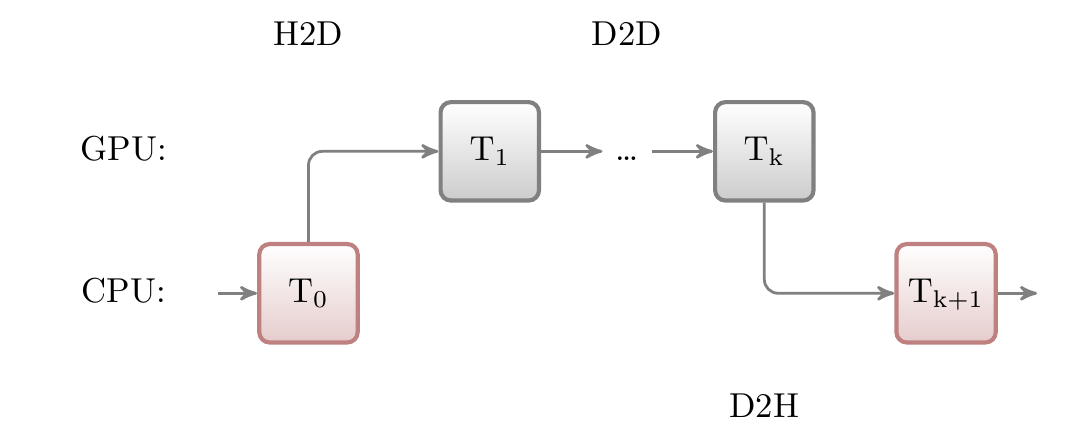}
  \vspace{-\figureC}
  \caption{An example of hybrid CPU/GPU graph.
    %Types of the data transfer: H2D --- Host to Device, D2D --- Device to Device, D2H --- Device to Host.
  }
  \label{fig:transfers}
  \vspace{-\figureB}
\end{figure}

The computational chain example with GPU transformations involved is shown in figure~\ref{fig:transfers}. A series of
GPU based transformations $T_1$ to $T_k$ are situated between CPU based transformations $T_0$ and $T_{k+1}$. A memory
transfer from host to device (H2D) is invoked when a $T_1$ transformation is called, a transfer back from device to host
(D2H) is triggered only when CPU based $T_{k+1}$ reads the output of GPU based $T_k$. It is essential that no data
transfer between host and device is invoked between consequential GPU transformations.

The overhead on the GPU implementation call and memory synchronization should be taken into account.
Therefore, the number of data transfers and function calls should be minimized. On a framework level this will be done by
incorporating individual data transfers into a single one. On a user level the computational chain should be designed
appropriately: the bigger and more continuous portion of the graph being switched to the GPU the better.
It is more appropriate when the inputs to the GPU subgraph do not depend on any parameters.
Also, in contrast to the CPU version, when high granularity of the chain is desired in order to profit from caching and
lazy evaluation, the GPU version should be clustered into a narrow graph to maximize the amount of data per single GPU
transformation call. Tools will be provided for effortless switching between these schemes.

It is also worth noting that typical GPUs provide superior efficiency for single precision floats when
compared to double precision. A way to choose the computational chain precision will be provided so that the user can check
the model performance and switch to single precision.

Further prospects of combined CPU/GPU usage are also considered. They include the hybrid case when different GPU
transformations are invoked from parallel threads. In the massively parallel case several dozens of replicas of the same
model may be merged and executed on GPU in a single call. Therefore the maximal profit may be achieved in a case when
these models are used for parallel fitting to different MC samples or different points of a studied parameter space.

\section{Conclusion}

We present an overview of methods implemented in GNA, a framework for statistical analysis of large-scale models. GNA
uses the data flow paradigm to provide a way to build caching and efficient lazily evaluated models. The models may be
assembled at runtime, thus giving the user high flexibility.
There are inspiring prospects to include, implicitly from the end user point of view, multithreading on CPU and GPU support.

The framework is currently under active development with many features introduced recently. As long as the code base
is not yet finalized the development is carried out in a private repository.
After finalizing and cross checking the Daya Bay~\cite{Gonchar:1810db} model, we will suspend the active development
phase in order to prepare a production release. We plan to review and document the GNA framework in early 2019 and make the
project public by Spring 2019.

The current project page may be found in~\cite{gna:astronu}. The documentation page~\cite{gna:pages} is slightly
outdated while the reference guide section~\cite{gna:refguide} contains an actual description of the subset of
transformations and tools.

\section*{Acknowledgements}

The research is supported by the Russian Foundation for Basic Research
(grant 18-32-00935) and by Association of Young Scientists and
Specialists of JINR (grant 18-202-08).

We are grateful to Chris Kullenberg for reading the manuscript and for the valuable comments.

\end{document}